\documentstyle[times,twocolumn,latex8]{article}

\newcommand{\st}[1]{|{#1}\rangle} 
\newcommand{\ceil}[1]{\lceil{#1}\rceil} 
\newcommand{\R}{\mbox{\bf R}} 
\newcommand{\OR}{\mbox{\rm OR}}  
\def\01{\{0,1\}} 

\newtheorem{definition}{Definition}[section]
\newtheorem{theorem}[definition]{Theorem}
\newtheorem{proposition}[definition]{Proposition}
\newtheorem{lemma}[definition]{Lemma}
\newtheorem{corollary}[definition]{Corollary}

\newenvironment{proof}
{\noindent {\bf Proof }}
{{\hfill $\Box$}\\
}

\begin{document}

\bibliographystyle{latex8}

\title{Quantum Lower Bounds by Polynomials\thanks{Part of this work was 
done while the third and fourth authors were visiting CWI in December 1997.}}
\author{\addtocounter{footnote}{1}
Robert Beals\\
{\protect\small\sl University of Arizona\/}%
\thanks{Department of Mathematics, University of Arizona, P.O.~Box 210089,
617 N.\ Santa Rita Ave, Tucson AZ 85721--0089, USA. 
E-mail: {\tt beals@math.arizona.edu}.}
\and
Harry Buhrman\\
{\protect\small\sl CWI, Amsterdam\/}%
\thanks{CWI, P.O.~Box 94079, Amsterdam, The Netherlands. 
E-mail: {\tt buhrman@cwi.nl}.}
\and
Richard Cleve\\
{\protect\small\sl University of Calgary\/}
\thanks{Department of Computer Science, University of Calgary, Calgary, Alberta,
Canada T2N 1N4. E-mail: {\tt cleve@cpsc.ucalgary.ca}.}
\and
Michele Mosca\\
{\protect\small\sl University of Oxford\/}
\thanks{Mathematical Institute, University of Oxford, 24-29 St.~Giles', 
Oxford,OX1 3LB, U.K.,
and Centre for Quantum Computation,
Clarendon Laboratory, Parks Road, Oxford, OX1 3PU, U.K.
E-mail: {\tt mosca@maths.ox.ac.uk}.}
\and
Ronald de Wolf\\
{\protect\small\sl CWI and University of Amsterdam\/}
\thanks{CWI, P.O.~Box 94079, Amsterdam, The Netherlands. E-mail: {\tt rdewolf@cwi.nl}.}
}
\maketitle
\thispagestyle{empty}

\begin{abstract}
We examine the number $T$ of queries that a quantum network requires 
to compute several Boolean functions on $\{0,1\}^N$ in the {\em black-box} 
model.
We show that, in the black-box model, the exponential quantum speed-up 
obtained for {\em partial\/} functions (i.e.\ problems involving a promise
on the input) by Deutsch and Jozsa and by Simon cannot be obtained 
for any {\em total\/} function: 
if a quantum algorithm computes some total Boolean function $f$ 
with bounded-error using $T$ black-box queries then there is a classical 
deterministic algorithm that computes $f$ exactly with $O(T^6)$ queries.
We also give asymptotically tight characterizations of $T$ for all 
symmetric $f$ in the exact, zero-error, and bounded-error settings.
Finally, we give new precise bounds for AND, OR, and PARITY.
Our results are a quantum extension of the so-called polynomial method, 
which has been successfully applied in classical complexity theory, 
and also a quantum extension of results by Nisan about a polynomial 
relationship between randomized and deterministic decision tree complexity.
\end{abstract}

\section{Introduction}

The {\em black-box} model of computation arises when one is given a
black-box containing an $N$-tuple of Boolean variables
$X=(x_0,x_1,\ldots,x_{N-1})$.  The box is equipped to output $x_i$ on
input $i$. We wish to determine some property of $X$, accessing
the $x_i$ only through the black-box.  
Such a black-box access is called a {\em query}.
A property of $X$ is any Boolean function that depends on $X$,
i.e.\ a property is a function $f : \01^N \rightarrow \01$.
We want to compute such properties using as few queries as possible.

Consider, for example, the case where the goal is to determine whether 
or not $X$ contains at least one 1, so we want to compute the property
$\OR(X) = x_0\vee\ldots\vee x_{N-1}$.
It is well known that the number of queries required to compute $\OR$ 
by any {\em classical\/} (deterministic or probabilistic) algorithm 
is $\Theta(N)$.
Grover~\cite{grover:search} discovered a remarkable {\em quantum} algorithm 
that, making queries in superposition, can be used to compute $\OR$ with 
small error probability using only $O(\sqrt{N})$ queries.
This number of queries was shown to be asymptotically 
optimal~\cite{bbbv:str&weak,bbht:bounds,zalka:grover}.

Many other quantum algorithms can be naturally expressed in the black-box 
model, such as an algorithm due to Simon~\cite{simon:power}, in which 
one is given a function $\tilde{X} : \01^n \rightarrow \01^n$, 
which, technically, can also be viewed as a black-box
$X=(x_0,\ldots,x_{N-1})$ with $N = n 2^n$.
The black-box $X$ satisfies a particular promise, and the goal is to 
determine whether or not $X$ satisfies some other property (the details 
of the promise and properties are explained in~\cite{simon:power}).
Simon's quantum algorithm is proven to yield an {\em exponential\/} 
speed-up over classical algorithms in that it makes $(\log N)^{O(1)}$ queries, 
whereas every classical randomized algorithm for the same function must 
make $N^{\Omega(1)}$ queries.
The promise means that the function $f : \01^N \rightarrow \01$ is 
{\em partial\/}; it is not defined on all $X\in\01^N$.
(In the previous example of OR, the function is {\em total\/}; however, 
the quantum speed-up is only quadratic.)
Some other quantum algorithms that are naturally expressed in the black-box 
model are described in~\cite{deutsch&jozsa,boneh&lipton,kitaev:stabilizer,bbht:bounds,brassard&hoyer:simon,hoyer:conjugated,mosca&ekert:hidden,cemm:revisited,bht:counting,mosca:eigen,BuhrmanCleveWigderson98}.

Of course, {\em upper bounds} in the black-box model immediately yield 
upper bounds for the {\em circuit description\/} model in which the function 
$X$ is succinctly described as a $(\log N)^{O(1)}$-sized circuit computing 
$x_i$ from $i$.
On the other hand, {\em lower bounds} in the black-box model do not imply 
lower bounds in the circuit model, though they can provide useful guidance, 
indicating what certain algorithmic approaches are capable of accomplishing.
It is noteworthy that, at present, there is no known algorithm for computing 
$\OR$ (i.e.\ satisfiability) in the circuit model that is significantly more 
efficient than using the circuit solely to make queries (though, {\em proving} 
that no better algorithm exists is likely to be difficult, as it would 
imply $P \neq NP$).

It should also be noted that the black-box complexity of a function 
only considers the number of queries; it does
not capture the complexity of the {\em auxiliary} computational steps
that have to be performed in addition to the queries. In cases such
as OR, PARITY, MAJORITY, this auxiliary work is not significantly
larger than the number of queries; however, in some cases it may be
much larger. For example, consider the case of factoring N-bit
integers. The best known algorithms for this involve $\Theta(N)$ queries
to determine the integer, followed by $2^{N^{\Omega(1)}}$ operations 
in the classical case but only $N^2(\log N)^{O(1)}$ operations in the
quantum case~\cite{shor:factoring}. Thus, the number of queries is 
apparently not of primary importance in the case of factoring.

In this paper, we analyze the black-box complexity of several functions 
and classes of functions in the quantum computation setting.
In particular, we show that the kind of exponential quantum speed-up that 
Simon's algorithm achieves for a partial function cannot be obtained by 
any quantum algorithm for any total function: at most a polynomial
speed-up is possible.
We also tightly characterize the quantum black-box complexity of all 
symmetric functions, and obtain exact bounds for functions such as AND, OR, 
PARITY, and MAJORITY for various error models: exact, zero-error, bounded-error.

An important ingredient of our approach is a reduction that translates 
quantum algorithms that make $T$ queries into  multilinear polynomials 
over the $N$ variables of degree at most $2T$.
This is a quantum extension of the so-called polynomial method, which has 
been successfully applied in classical complexity theory 
(see~\cite{beigel:poly} for an overview).
Also, our polynomial relationship between the quantum and the 
classical complexity is analogous to earlier results by 
Nisan~\cite{nisan:pram&dt}, who proved a polynomial relationship between 
randomized and deterministic decision tree complexity.

\section{Summary of results}

We consider three different settings for computing $f$ on $\{0,1\}^N$ in 
the black-box model. 
In the {\em exact\/} setting, an algorithm is required to return $f(X)$ 
with certainty for every $X$.
In the {\em zero-error} setting, for every $X$, an algorithm may return 
``inconclusive'' with probability at most $1/2$, but {\em if\/} it returns 
an answer, this must be the correct value of $f(X)$ (algorithms in this 
setting are sometimes called {\em Las Vegas} algorithms).
Finally, in the {\em two-sided bounded-error} setting, for every $X$, 
an algorithm must correctly return the answer with probability at 
least $2/3$ (algorithms in this setting are sometimes called 
{\em Monte Carlo} algorithms; the $2/3$ is arbitrary).
Our main results are:%
\footnote{All our results remain valid if we consider a {\em controlled}
black-box, where the first bit of the state indicates whether the black-box is
to be applied or not. (Thus such a black-box would map $\st{0,i,b,z}$ to
$\st{0,i,b,z}$ and $\st{1,i,b,z}$ to $\st{1,i,b\oplus x_i,z}$.)
Also, our results remain valid if we consider {\em mixed} 
rather than only pure states.}
\begin{enumerate}
\item
In the black-box model, the quantum speed-up for {\em any} total function 
cannot be more than by a sixth-root.
More specifically, if a quantum algorithm computes $f$ with bounded-error 
probability by making $T$ queries, then there is a classical deterministic 
algorithm that computes $f$ exactly making at most $O(T^6)$ queries.
If $f$ is {\em monotone} then the classical algorithm needs at most $O(T^4)$ 
queries, and if $f$ is {\em symmetric} then it needs at 
most $O(T^2)$ queries.

As a by-product, we also improve the polynomial relation between 
the {\em decision tree complexity $D(f)$} and the {\em approximate degree}
$\widetilde{deg}(f)$ of~\cite{nisan&szegedy:degree}
from $D(f)\in O(\widetilde{deg}(f)^8)$ to $D(f)\in O(\widetilde{deg}(f)^6)$.
\item
We tightly characterize the black-box complexity of all non-constant 
symmetric functions as follows.
In the exact or zero-error settings $\Theta(N)$ queries are necessary and sufficient, 
and in the bounded-error setting $\Theta(\sqrt{N(N-\Gamma(f))})$ queries
are necessary and sufficient, 
where $\Gamma(f)=\min\{|2k-N+1| :$ $f$ flips value if the Hamming weight of
the input changes from $k$ to $k+1\}$ (this $\Gamma(f)$ is a number that is low if $f$ 
flips for inputs with Hamming weight close to $N/2$~\cite{paturi:degree}).
This should be compared with the {\em classical\/} bounded-error query 
complexity of such functions, which is $\Theta(N)$.
Thus, $\Gamma(f)$ characterizes the speed-up that quantum algorithms give.

An interesting example is the THRESHOLD$_{M}$ function which is 1 iff 
its input $X$ contains at least $M$ 1s.  
This has query complexity $\Theta(\sqrt{M(N-M+1)})$.
\item
For OR, AND, PARITY, MAJORITY, we obtain the bounds in the table below 
(all given numbers are both necessary and sufficient).
\begin{table}[htb]
\centering
\begin{tabular}{|l|c|c|c|c|}
\hline
 & exact & zero-error & bounded-error\\ \hline\hline
OR, AND  & $N$ & $N$ & $\Theta(\sqrt{N})$ \\ \hline
PARITY   & $N/2$ & $N/2$ & $N/2$\\ \hline
MAJORITY & $\Theta(N)$ & $\Theta(N)$ & $\Theta(N)$\\ \hline
\end{tabular}
\caption{Some quantum complexities}
\end{table}
These results are all new, with the exception of the 
$\Theta(\sqrt{N})$-bounds for OR and AND in the bounded-error setting, 
which appear in~\cite{grover:search,bbbv:str&weak,bbht:bounds,zalka:grover}.
The new bounds improve by polylog($N$) factors previous lower bound 
results from~\cite{BuhrmanCleveWigderson98}, 
which were obtained through a reduction from communication complexity.
The new bounds for PARITY were independently obtained by Farhi 
{\it et al.}~\cite{fggs:parity}.

Note that lower bounds for OR imply lower bounds for {\em database search} 
(where we want to find an $i$ such that $x_i=1$, if one exists), so exact 
or zero-error quantum search requires $N$ queries, in contrast 
to $\Theta(\sqrt{N})$ queries for the bounded-error case.
\end{enumerate}

\section{Preliminaries}

Our main goal in this paper is to find the number of queries a quantum algorithm
needs to compute some Boolean function by relating such networks to polynomials.
In this section we give some basic definitions and properties of multilinear
polynomials and Boolean functions, and describe our quantum setting.

\subsection{Boolean functions and polynomials}

We assume the following setting, mainly adapted from~\cite{nisan&szegedy:degree}.
We have a vector of $N$ Boolean variables $X=(x_0,\ldots,x_{N-1})$,
and we want to compute a Boolean function $f: \{0,1\}^N\rightarrow\{0,1\}$ of $X$.
Unless explicitly stated otherwise, $f$ will always be total.
The Hamming weight (number of 1s) of $X$ is denoted by $|X|$.
For convenience we will assume $N$ even, unless explicitly stated otherwise.
We can represent Boolean functions using $N$-variate polynomials $p: \R^N\rightarrow\R$. 
Since $x^k=x$ whenever $x\in\{0,1\}$, 
we can restrict attention to {\em multilinear} $p$.
If $p(X)=f(X)$ for all $X\in\{0,1\}^N$, then we say $p$ {\em represents} $f$.
We use $deg(f)$ to denote the degree of a minimum-degree $p$ that represents
$f$ (actually such a $p$ is unique).
If $|p(X)-f(X)|\leq 1/3$ for all $X\in\{0,1\}^N$, we say $p$ 
{\em approximates} $f$, and $\widetilde{deg}(f)$ denotes the degree of a minimum-degree 
$p$ that approximates $f$.
For example, $x_0x_1\ldots x_{N-1}$ is a multilinear polynomial of degree $N$ 
that represents the AND-function.
Similarly, $1-(1-x_0)(1-x_1)\ldots(1-x_{N-1})$ represents OR.
The polynomial $\frac{1}{3}x_0+\frac{1}{3}x_1$ approximates but does not represent
AND on 2 variables.

Nisan and Szegedy~\cite[Theorem~2.1]{nisan&szegedy:degree} proved 
a general lower bound on the degree of any Boolean function that depends 
on $N$ variables:

\begin{theorem}[Nisan, Szegedy]\label{thdeglogn}
If $f$ is a Boolean function that depends on $N$ variables,
then $deg(f)\geq \log N-O(\log\log N)$.
\end{theorem}

Let $p: \R^N\rightarrow\R$ be a polynomial.
If $\pi$ is some permutation and $X=(x_0,\ldots,x_{N-1})$, 
then $\pi(X)=(x_{\pi(0)},\ldots,x_{\pi(N-1)})$.
Let $S_N$ be the set of all $N!$ permutations. 
The {\em symmetrization} $p^{sym}$ of $p$ averages over all permutations of the
input, and is defined as:
$$
p^{sym}(X)=
\frac{\sum_{\pi\in S_N}p(\pi(X))}{N!}.
$$
Note that $p^{sym}$ is a polynomial of degree at most the degree of $p$.
Symmetrizing may actually lower the degree: if $p=x_0-x_1$, then $p^{sym}=0$.
The following lemma, originally due to~\cite{minsky&papert:perceptrons},
allows us to reduce an $N$-variate polynomial to a single-variate one.

\begin{lemma}[Minsky, Papert]\label{lemsym}
If $p: \R^n\rightarrow\R$ is a multilinear polynomial, then there exists a
polynomial $q: \R\rightarrow\R$, of degree at most the degree of $p$,
such that $p^{sym}(X)=q(|X|)$ for all $X\in\{0,1\}^N$.
\end{lemma}

\begin{proof}
Let $d$ be the degree of $p^{sym}$, which is at most the degree of $p$.
Let $V_j$ denote the sum of all $N\choose j$ products of $j$ different
variables, so $V_1=x_0+\ldots+x_{N-1}$,
$V_2=x_0x_1+x_0x_2+\ldots+x_{N-1}x_{N-2}$, etc.
Since $p^{sym}$ is symmetrical, it can be written as
$$
p^{sym}(X)=a_0+a_1V_1+a_2V_2+\ldots+a_dV_d,
$$
for some $a_i\in\R$.
Note that $V_j$ assumes value
${|X|\choose j}=|X|(|X|-1)(|X|-2)\ldots(|X|-j+1)/j!$ on $X$,
which is a polynomial of degree $j$ of $|X|$.
Therefore the single-variate polynomial $q$ defined by
$$
q(|X|)=a_0+a_1{|X|\choose 1}+a_2{|X|\choose 2}+\ldots+a_d{|X|\choose d}
$$
satisfies the lemma.
\end{proof}

A Boolean function $f$ is {\em symmetric} if permuting the input does not
change the function value (i.e.,\ $f(X)$ only depends on $|X|$).
Paturi has proved a powerful theorem that characterizes $\widetilde{deg}(f)$
for symmetric $f$.
For such $f$, let $f_k=f(X)$ for $|X|=k$, and define 
$$
\Gamma(f)=\min\{|2k-N+1| : f_k\neq f_{k+1} \mbox{ and } 0\leq k\leq N-1\}.
$$
$\Gamma(f)$ is low if $f_k$ ``jumps'' near the middle (i.e.,\ for some 
$k\approx N/2$). Now~\cite[Theorem~1]{paturi:degree} gives:

\begin{theorem}[Paturi]\label{thappdegreesym}
If $f$ is a non-constant symmetric Boolean function on $\{0,1\}^N$, 
then $\widetilde{deg}(f)\in\Theta(\sqrt{N(N-\Gamma(f))})$.
\end{theorem}

For functions like OR and AND, we have $\Gamma(f)=N-1$ and hence 
$\widetilde{deg}(f)\in\Theta(\sqrt{N})$.
For PARITY (which is 1 iff $|X|$ is odd) and 
MAJORITY (which is 1 iff $|X|>N/2$),
we have $\Gamma(f)=1$ and $\widetilde{deg}(f)\in\Theta(N)$.

\subsection{The framework of quantum networks}

Our goal is to compute some Boolean function $f$ of $X=(x_0,\ldots,x_{N-1})$,
where $X$ is given as a black-box: calling the black-box on $i$ returns 
the value of $x_i$. We want to use as few queries as possible.

A classical algorithm that computes $f$ by using (adaptive)
black-box queries to $X$ is called a {\em decision tree}, 
since it can be pictured as a binary tree
where each node is a query, each node has the two outcomes of the query 
as children, and the leaves give answer $f(X)=0$ or $f(X)=1$.
The {\em cost} of such an algorithm is the number of queries made on the
worst-case $X$, so the cost is the depth of the tree.
The {\em decision tree complexity} $D(f)$ of $f$ is the cost of the best  
decision tree that computes $f$.
Similarly we can define $R(f)$ as the expected number of queries on 
the worst-case $X$ for
{\em randomized} algorithms that compute $f$ with bounded-error.

A {\em quantum network} with $T$ queries is the quantum analogue to 
a classical decision tree with $T$ queries, where queries and other 
operations can now be made in quantum superposition.
Such a network can be represented as a sequence of unitary transformations:
$$
U_0,O_1,U_1,O_2,\ldots,U_{T-1},O_T,U_T,
$$
where the $U_i$ are arbitrary unitary transformations, and the $O_j$
are unitary transformations which correspond to queries to $X$.
The computation ends with some measurement or observation of the final state.
We assume each transformation acts on $m$ qubits and each qubit has basis
states $\st{0}$ and $\st{1}$, so there are $2^m$ basis states for each
stage of the computation. 
It will be convenient to represent each basis state as a binary string of 
length $m$ or as the corresponding natural number, so we have basis states 
$\st{0},\st{1},\st{2},\ldots,\st{2^m-1}$. 
Let $K$ be the index set $\{0,1,2,\ldots,2^m-1\}$.
With some abuse of notation, we will sometimes identify a set of numbers
with the corresponding set of basis states.
Every state $\st{\phi}$ of the network can be uniquely written as 
$\st{\phi}=\sum_{k\in K}\alpha_k\st{k}$, where the $\alpha_k$ are complex 
numbers such that $\sum_{k\in K}|\alpha_k|^2=1$.
When $\st{\phi}$ is measured in the above basis, 
the probability of observing $\st{k}$ is $|\alpha_k|^2$.
Since we want to compute a function of $X$, which is given as a black-box, 
the initial state of the network is not very important and we will disregard 
it hereafter (we may assume the initial state to be $\st{0}$ always).

The queries are implemented using the unitary transformations $O_j$
in the following standard way. 
The transformation $O_j$ only affects the leftmost part of a basis state:
it maps basis state $\st{i,b,z}$ to $\st{i,b\oplus x_i,z}$ ($\oplus$ denotes
XOR).
Here $i$ has length $\ceil{\log N}$ bits, $b$ is one bit, and $z$ is an
arbitrary string of $m-\ceil{\log N}-1$ bits. 
Note that the $O_j$ are all equal.

How does a quantum network compute a Boolean function $f$ of $X$?
Let us designate the rightmost bit of the final state of the network as the output bit. 
More precisely, the output of the computation is defined to be 
the value we observe if we measure the rightmost bit of the final state.
If this output equals $f(X)$ with certainty, for every $X$, 
then the network computes $f$ {\em exactly}. 
If the output equals $f(X)$ with probability at least $2/3$, for every $X$, 
then the network computes $f$ with bounded error probability at most $1/3$.
To define the zero-error setting, the output is obtained by observing the {\em two}
rightmost bits of the final state. If the first of these bits is 0, the network
claims ignorance (``inconclusive''), otherwise the second bit should contain $f(X)$
with certainty.
For every $X$, the probability of getting ``inconclusive'' should be less than $1/2$.
We use $Q_E(f)$, $Q_0(f)$ and $Q_2(f)$ to denote the minimum number of queries
required by a quantum network to compute $f$ in the exact, 
zero-error and bounded-error settings, respectively.
Note that $Q_2(f)\leq Q_0(f)\leq Q_E(f)\leq D(f)\leq N$.

\section{General lower bounds on the number of queries}\label{secbounds}

In this section we will provide some general lower bounds on the number 
of queries required to compute a Boolean function $f$ on a quantum network, 
either exactly or with zero- or bounded-error probability.

\subsection{Bounds for error-free computation}

The next lemmas relate quantum networks to polynomials; they are the key
to most of our results.

\begin{lemma}\label{lemamplpol}
Let $\cal N$ be a quantum network that makes $T$ queries to a black-box $X$. 
Then there exist complex-valued $N$-variate multilinear polynomials 
$p_0,\ldots,p_{2^m-1}$, each of degree at most $T$, such that the final state
of the network is the superposition 
$$
\sum_{k\in K}p_k(X)\st{k},
$$
for any black-box $X$.
\end{lemma}

\begin{proof}
Let $\st{\phi_i}$ be the state of the network (using some black-box $X$)
just before the $i$th query. 
Note that $\st{\phi_{i+1}}=U_iO_i\st{\phi_i}$.
The amplitudes in $\st{\phi_0}$ depend on the initial state and on
$U_0$ but not on $X$, so they are polynomials of $X$ of degree 0. 
A query maps basis state $\st{i,b,z}$ to $\st{i,b\oplus x_i,z}$.
Hence if the amplitude of $\st{i,0,z}$ in $\st{\phi_0}$ is $\alpha$ and
the amplitude of $\st{i,1,z}$ is $\beta$, then 
the amplitude of $\st{i,0,z}$ {\em after} the query becomes
$(1-x_i)\alpha+x_i\beta$ and the amplitude of $\st{i,1,z}$ becomes
$x_i\alpha+(1-x_i)\beta$, which are polynomials of degree $1$.
(In general, if the amplitudes before a query are polynomials of 
degree $\leq j$, then the amplitudes after the query will be 
polynomials of degree $\leq j+1$.)
Between the first and the second query lies the unitary transformation
$U_1$. However, the amplitudes after applying $U_1$ are just linear 
combinations of the amplitudes before applying $U_1$, so the amplitudes
in $\st{\phi_1}$ are polynomials of degree at most $1$.
Continuing in this manner, the amplitudes of the final states are found
to be polynomials of degree at most $T$. 
We can make these polynomials multilinear without affecting their values 
on $X\in\01^N$, by replacing all $x_i^k$ by $x_i$.
\end{proof}

Note that we have not used the assumption that the $U_j$ are unitary,
but only their linearity.
The next lemma is also implicit in the combination of some proofs
in~\cite{ffkl:toolkit,fortnow&rogers:limitations}.

\begin{lemma}\label{lemprobpol}
Let $\cal N$ be a quantum network that makes $T$ queries to a black-box $X$,
and $B$ be a set of basis states. 
Then there exists a real-valued multilinear polynomial $P(X)$ of degree at most $2T$, 
which equals the probability that observing the final state of the network 
with black-box $X$ yields a state from $B$.
\end{lemma}

\begin{proof}
By the previous lemma, we can write the final state of the network as
$$
\sum_{k\in K}p_k(X)\st{k},
$$
for any $X$, where the $p_k$ are complex-valued 
polynomials of degree $\leq T$.
The probability of observing a state in $B$ is
$$
P(X)=\sum_{k\in B}|p_k(X)|^2.
$$
If we split $p_k$ into its real and imaginary parts as 
$p_k(X)=pr_k(X)+i\cdot pi_k(X)$, where $pr_k$ and $pi_k$ are real-valued 
polynomials of degree $\leq T$, then $|p_k(X)|^2=(pr_k(X))^2+(pi_k(X))^2$,
which is a real-valued polynomial of degree at most $2T$.
Hence $P$ is also a real-valued polynomial of degree at most $2T$,
which we can make multilinear without affecting its values on $X\in\01^N$.
\end{proof}

Letting $B$ be the set of states that have 1 as rightmost bit, 
it follows that we can write the acceptance probability of a network as a degree-$2T$
polynomial $P(X)$ of $X$. In the case of exact computation of $f$ we must have
$P(X)=f(X)$ for all $X$, so $P$ represents $f$ and we obtain $2T\geq deg(f)$.

\begin{theorem}\label{thexactpol}
If $f$ is a Boolean function, then $Q_E(f)\geq deg(f)/2$.
\end{theorem}

Combining this with Theorem~\ref{thdeglogn}, we obtain a general lower bound:

\begin{corollary}
If $f$ depends on $N$ variables,
then $Q_E(f)\geq (\log N)/2-O(\log\log N)$.
\end{corollary}

For {\em symmetric} $f$ we can prove a much stronger bound.
Firstly for the zero-error setting:

\begin{theorem}\label{thsymzero}
If $f$ is non-constant and symmetric,
then $Q_0(f)\geq (N+1)/4$.
\end{theorem}

\begin{proof}
We assume $f(X)=0$ for at least $(N+1)/2$ different Hamming weights of $X$; 
the proof is similar if $f(X)=1$ for at least $(N+1)/2$ different Hamming weights.
Consider a network that uses $T=Q_0(f)$ queries to compute $f$ with zero-error.
Let $B$ be the set of basis states that have $11$ as rightmost bits.
By Lemma~\ref{lemprobpol}, there is a real-valued multilinear polynomial $P$ 
of degree $\leq 2T$, such that for all $X$, $P(X)$ equals the probability 
that the output of the network is $11$ (i.e.,\ that the network answers 1). 
Since the network computes $f$ with zero-error and $f$ is non-constant, $P(X)$
is non-constant and equals 0 on at least $(N+1)/2$ different Hamming weights
(namely the Hamming weights for which $f(X)=0$).
Let $q$ be the single-variate polynomial of degree $\leq 2T$
obtained from symmetrizing $P$ (Lemma~\ref{lemsym}).
This $q$ is non-constant and has at least $(N+1)/2$ zeroes, hence degree
at least $(N+1)/2$, and the result follows.
\end{proof}

Thus functions like OR, AND, PARITY, threshold functions  etc., 
all require at least $(N+1)/4$ queries to be computed exactly or with
zero-error on a quantum network.
Since $N$ queries always suffice, even classically, we have $Q_E(f)\in\Theta(N)$ 
and $Q_0(f)\in\Theta(N)$ for non-constant symmetric $f$.

Secondly, for the exact setting, we can use results by
Von zur Gathen and Roche~\cite[Theorems~2.6 and~2.8]{gathen&roche:poly}:

\begin{theorem}[Von zur Gathen, Roche]
If $f$ is non-constant and symmetric, then $deg(f)=N-O(N^{0.548})$.
If, in addition, $N+1$ is prime, then $deg(f)=N$.
\end{theorem}

\begin{corollary}
If $f$ is non-constant and symmetric, then $Q_E(f)\geq N/2 - O(N^{0.548})$.
If, in addition, $N+1$ is prime, then $Q_E(f)\geq N/2$.
\end{corollary}

In Section~\ref{secpartfns} we give more precise bounds for some particular 
functions. In particular, this will show that the $N/2$ lower bound is tight,
as it can be met for PARITY.

\subsection{Bounds for computation with bounded-error}\label{ssecappr}

Here we use similar techniques to get bounds on the number of queries
required for {\em bounded-error} computation of some function.
Consider the acceptance probability of a $T$-query network that computes $f$
with bounded-error, written as a polynomial $P(X)$ of degree $\leq 2T$. 
If $f(X)=0$ then we should have $P(X)\leq 1/3$, and if $f(X)=1$ then $P(X)\geq 2/3$.
Hence $P$ approximates $f$, and we get:

\begin{theorem}\label{thapprpol}
If $f$ is a Boolean function,
then $Q_2(f)\geq\widetilde{deg}(f)/2$.
\end{theorem}

This result implies that a quantum algorithm that computes $f$
with bounded error probability can be at most polynomially more efficient 
(in terms of number of queries) than a classical deterministic algorithm:
Nisan and Szegedy proved that $D(f)\in O(\widetilde{deg}(f)^8)$
\cite[Theorem~3.9]{nisan&szegedy:degree}, which together with the previous 
theorem implies $D(f)\in O(Q_2(f)^8)$.
The fact that there is a polynomial relation between the classical and
the quantum complexity is also implicit in the generic oracle-constructions
of Fortnow and Rogers~\cite{fortnow&rogers:limitations}.
In Section~\ref{secpolrel} we will prove the stronger result
$D(f)\in O(Q_2(f)^6)$.

Combining Theorem~\ref{thapprpol} with Paturi's Theorem~\ref{thappdegreesym}
gives a lower bound for {\em symmetric} functions 
in the bounded-error setting: if $f$ is non-constant and symmetric, 
then $Q_2(f)=\Omega(\sqrt{N(N-\Gamma(f))})$.
We can in fact prove a matching upper bound, using the following result,
which follows immediately from~\cite{bht:counting} 
as noted by Mosca~\cite{mosca:eigen}. It shows that we can {\em count} 
the number of 1s in $X$ exactly, with bounded error probability:

\begin{theorem}[Brassard, H{\o}yer, Tapp; Mosca]\label{thqcount}
There exists a quantum algorithm that returns $t=|X|$ with probability
at least $3/4$ using expected time $\Theta(\sqrt{(t+1)(N-t+1)})$, 
for all $X\in\{0,1\}^N$.
\end{theorem}

Actually, the algorithms given in \cite{bht:counting, mosca:eigen} are
classical algorithms which use some quantum networks as subroutines; the
notion of {\em expected} time for such algorithms is the same as for classical
ones.
This counting-result allows us to prove the matching upper bound:

\begin{theorem}\label{thcountsym}
If $f$ is non-constant and symmetric,
then $Q_2(f)\in\Theta(\sqrt{N(N-\Gamma(f))})$.
\end{theorem}

\begin{proof}
Let $f$ be some non-constant Boolean function.
We will sketch a strategy that computes $f$ with bounded error 
probability $\leq 1/3$. 
Let $f_k=f(X)$ for $X$ with $|X|=k$.
First note that since 
$\Gamma(f)=\min\{|2k-N+1| : f_k\neq f_{k+1} \mbox{ and } 0\leq k\leq N-1\}$,
$f_k$ must be identically 0 or 1 for 
$k\in\{(N-\Gamma(f))/2,\ldots,(N+\Gamma(f)-2)/2\}$.  
Consider some $X$ with $|X|=t$.
In order to be able to compute $f(X)$, it is sufficient to know $t$ exactly 
if $t<(N-\Gamma(f))/2$ or $t>(N+\Gamma(f)-2)/2$, or to {\em know} that
$(N-\Gamma(f))/2\leq t\leq(N+\Gamma(f)-2)/2$ otherwise.

Run the counting algorithm for $\Theta(\sqrt{(N-\Gamma(f))N/2})$
steps to count the number of 1s in $X$. 
If $t<(N-\Gamma(f))/2$ or $t>(N+\Gamma(f)-2)/2$, 
then with high probability the algorithm
will have terminated and will have returned $t$. 
If it has not terminated after $\Theta(\sqrt{(N-\Gamma(f))N/2})$ 
steps, then we know $(N-\Gamma(f))/2 \leq t \leq (N+\Gamma(f)-2)/2$ 
with high probability.

 From this application of the counting algorithm,
we now have obtained the following with bounded error probability: 
\begin{itemize}
\item If $t<(N-\Gamma(f))/2$ or $t>(N+\Gamma(f)-2)/2$,
then the counting algorithm gave us an exact count of $t$.
\item If $(N-\Gamma(f))/2\leq t\leq(N+\Gamma(f)-2)/2$, then we know this, and we
also know that $f_t$ is identically 0 or 1 for all such $t$.
\end{itemize}
Thus with bounded error probability we have obtained sufficient information 
to compute $f_t=f(X)$, using only $O(\sqrt{N(N-\Gamma(f))})$ queries.
Repeating this procedure some constant number of times, we can limit 
the probability of error to at most $1/3$.
We can implement this strategy in a quantum network with 
$O(\sqrt{N(N-\Gamma(f))})$ 
queries to compute $f$.
\end{proof}

This implies that the above-stated result about quantum counting (Theorem~\ref{thqcount})
is optimal, since a better upper bound for counting would give a better upper 
bound on $Q_2(f)$ for symmetric $f$, 
whereas we already know that Theorem~\ref{thcountsym} is tight.
In contrast to Theorem~\ref{thcountsym}, it can be shown that a randomized classical strategy needs
$\Theta(N)$ queries to compute any non-constant symmetric $f$ with bounded-error.

After reading a first version of this paper, where we proved that 
most functions cannot be computed exactly using significantly fewer than $N$ 
(i.e.,\ $o(N)$) queries, Andris Ambainis~\cite{ambainis:comm}
extended this to the bounded-error case: {\em most} functions cannot 
be computed with bounded-error using significantly fewer than $N$ queries.

On the other hand, Wim van Dam~\cite{dam:oracle} recently proved
that with good probability we can learn all $N$ variables in the black-box
using only $N/2+\sqrt{N}$ queries. This implies the general upper bound
$Q_2(f)\leq N/2+\sqrt{N}$ for {\em any} $f$. This bound is almost tight, 
as we will show later on that $Q_2(f)=N/2$ for $f=$ PARITY.

\subsection{Lower bounds in terms of block sensitivity}\label{ssecbs}

Above we gave lower bounds on the number of queries used, in terms of degrees 
of polynomials that represent or approximate the function $f$ that is to be 
computed. Here we give lower bounds in terms of the {\em block sensitivity} of $f$.

\begin{definition}
Let $f: \{0,1\}^N\rightarrow\{0,1\}$ be a function, $X\in\{0,1\}^N$, 
and $B\subseteq\{0,\ldots,N-1\}$ a set of indices. 
Let $X^B$ denote the vector obtained from $X$ by flipping the variables in $B$.
We say that $f$ is {\em sensitive} to $B$ on $X$ if $f(X)\neq f(X^B)$.
The {\em block sensitivity} $bs_X(f)$ of $f$ {\em on $X$} is the maximum number 
$t$ for which there exist $t$ disjoint sets of indices $B_1,\ldots,B_t$ such 
that $f$ is sensitive to each $B_i$ on $X$.
The {\em block sensitivity} $bs(f)$ of $f$ is the maximum of $bs_X(f)$ 
over all $X\in\{0,1\}^N$.
\end{definition}

For example, $bs(\OR)=N$, because if we take $X=(0,0,\ldots,0)$ and
$B_i=\{i\}$, then flipping $B_i$ in $X$ flips the value of the
OR-function from 0 to 1.

We can adapt the proof of~\cite[Lemma~3.8]{nisan&szegedy:degree} on lower bounds
of polynomials to get lower bounds on the number of queries in a quantum network
in terms of block sensitivity.\footnote{This theorem can also be proved by 
an argument similar to the lower bound proof for database searching 
in~\cite{bbbv:str&weak}.}
The proof uses a theorem 
from~\cite{ehlich&zeller:schwankung,rivlin&cheney:approx}:

\begin{theorem}[Ehlich, Zeller; Rivlin, Cheney]
Let $p: \R\rightarrow\R$ be a polynomial such that
$b_1\leq p(i)\leq b_2$ for every integer $0\leq i\leq N$, and
$|p'(x)|\geq c$ for some real $0\leq x\leq N$.
Then $deg(p)\geq\sqrt{cN/(c+b_2-b_1)}$.
\end{theorem}

\begin{theorem}\label{thtbs}
If $f$ is a Boolean function,
then $Q_E(f)\geq\sqrt{bs(f)/8}$ and $Q_2(f)\geq\sqrt{bs(f)/16}$.
\end{theorem}

\begin{proof}
We will prove the theorem for bounded-error computation, 
the case of exact computation is completely analogous but slightly easier.
Consider a network using $T=Q_2(f)$ queries that computes $f$ with error probability $\leq 1/3$. 
Let $P$ be the polynomial of degree $\leq 2T$ that approximates $f$, obtained 
as for Theorem~\ref{thapprpol}.  Note that $P(X)\in[0,1]$ for all $X\in\{0,1\}^N$, 
because $P$ represents a probability.
Let $b=bs(f)$, and $X$ and $B_0,\ldots,B_{b-1}$ be the input and sets which 
achieve the block sensitivity. 
We assume without loss of generality that $f(X)=0$.

Consider variable $Y=(y_0,\ldots,y_{b-1})\in\R^b$.
Define $Z=(z_0,\ldots,z_{N-1})\in\R^N$ as:
$z_j=y_i$ if $x_j=0$ and $j\in B_i$,
$z_j=1-y_i$ if $x_j=1$ and $j\in B_i$,
and $z_j=x_j$ if $j\not\in B_i$ (the $x_j$ are fixed).
Note that if $Y=\vec{0}$ then $Z=X$, and if $Y$ has $y_i=1$ and $y_j=0$ for 
$j\neq i$ then $Z=X^{B_i}$.
Now $q(Y)=P(Z)$ is a $b$-variate polynomial of degree $\leq 2T$, such that
\begin{itemize}
\item $q(Y)\in[0,1]$ for all $Y\in\{0,1\}^b$ (because $P$ gives a probability).
\item $|q(\vec{0})-0|=|P(X)-f(X)|\leq 1/3$, so $0\leq q(\vec{0})\leq 1/3$. 
\item $|q(Y)-1|=|P(X^{B_i})-f(X^{B_i})|\leq 1/3$ if $Y$ has $y_i=1$ and $y_j=0$ 
for $j\neq i$.\\
Hence $2/3\leq q(Y)\leq 1$ if $|Y|=1$.
\end{itemize}
Let $r$ be the single-variate polynomial of degree $\leq 2T$
obtained from symmetrizing $q$ over $\{0,1\}^b$ (Lemma~\ref{lemsym}).
Note that $0\leq r(i)\leq 1$ for every integer $0\leq i\leq b$, and for 
some $x\in[0,1]$ we have $r'(x)\geq 1/3$ because $r(0)\leq 1/3$ and $r(1)\geq 2/3$.
Applying the previous theorem we get $deg(r)\geq\sqrt{b/4}$, hence $T\geq\sqrt{b/16}$.
\end{proof}

We can generalize this result to the computation of {\em partial}
Boolean functions, which only work on a domain ${\cal D}\subseteq\{0,1\}^N$
of inputs that satisfy some promise, by generalizing the definition of
block sensitivity to partial functions in the obvious way.

\section{Polynomial relation between classical and quantum complexity}\label{secpolrel}

Here we will compare the classical complexities $D(f)$ and $R(f)$ 
with the quantum complexities.
Some separations: 
as we show in the next section, if $f=$ PARITY then $Q_2(f)=N/2$ while $D(f)=N$;
if $f=$ OR then $Q_2(f)\in\Theta(\sqrt{N})$ by Grover's algorithm, while 
$R(f)\in\Theta(N)$ and $D(f)=N$, so we have a quadratic gap between
$Q_2(f)$ on the one hand and $R(f)$ and $D(f)$ on the other.%
\footnote{In the case of randomized decision trees, no function is known for 
which there is a quadratic gap between $D(f)$ and $R(f)$.
The best known separation is for complete binary AND/OR-trees, 
where $D(f)=N$ and $R(f)\in\Theta(N^{0.753\ldots})$, and it has been 
conjectured that this is the best separation possible. 
This holds both for zero-error randomized trees~\cite{saks&wigderson:trees} 
and for bounded-error trees~\cite{santha:montecarlo}.}

By a well-known result, the best randomized decision tree can be at most polynomially 
more efficient than the best deterministic decision tree: $D(f)\in O(R(f)^3)$
\cite[Theorem~4]{nisan:pram&dt}.
As mentioned in Section~\ref{secbounds}, we can prove that also the 
{\em quantum} complexity can be at most polynomially better than the best 
deterministic tree: $D(f)\in O(Q_2(f)^8)$. Here we give the stronger result that
$D(f)\in O(Q_2(f)^6)$. In other words, if we can compute some function quantumly with
bounded-error using $T$ queries, we can compute it classically error-free 
with $O(T^6)$ queries.

To start, we define the {\em certificate complexity} of $f$:

\begin{definition}
Let $f: \{0,1\}^N\rightarrow\{0,1\}$ be a function. 
A {\em $1$-certificate} is an assignment $C: S\rightarrow\{0,1\}$ of values 
to some subset $S$ of the $N$ variables, such that $f(X)=1$ whenever $X$ is 
consistent with $C$. The {\em size} of $C$ is $|S|$.
Similarly we define a $0$-certificate.

The {\em certificate complexity} $C_X(f)$ of $f$ {\em on $X$} is the size of 
a smallest $f(X)$-certificate that agrees with $X$.
The {\em certificate complexity} $C(f)$ of $f$ is the maximum of $C_X(f)$
over all $X$.
The {\em $1$-certificate complexity} $C^{(1)}(f)$ of $f$ is the maximum of 
$C_X(f)$ over all $X$ for which $f(X)=1$.
\end{definition}

For example, if $f$ is the OR-function, then the certificate complexity
on $(1,0,0,\ldots,0)$ is 1, because the assignment $x_0=1$ already forces
the OR to 1. The same holds for the other $X$ for which $f(X)=1$, so
$C^{(1)}(f)=1$. On the other hand, the certificate complexity on 
$(0,0,\ldots,0)$ is $N$, so $C(f)=N$.

The first inequality in the next lemma is obvious from the definitions,
the second inequality is \cite[Lemma~2.4]{nisan:pram&dt}.
We give the proof for completeness.

\begin{lemma}[Nisan]
$C^{(1)}(f)\leq C(f)\leq bs(f)^2$.
\end{lemma}

\begin{proof}
Consider an input $X\in\01^N$ and let $B_1,\ldots,B_b$ be disjoint {\em minimal} 
sets of variables that achieve the block sensitivity $b=bs_X(f)\leq bs(f)$.
We will show that $C:\cup_i B_i\rightarrow\{0,1\}$ which sets variables
according to $X$, is a certificate for $X$ of size $\leq bs(f)^2$.

Firstly, if $C$ were not an $f(X)$-certificate then let $X'$ be an input 
that agrees with $C$, such that $f(X')\neq f(X)$. Let $X'=X^{B_{b+1}}$.
Now $f$ is sensitive to $B_{b+1}$ on $X$ and $B_{b+1}$ is disjoint from 
$B_1,\ldots,B_b$, which contradicts $b=bs_X(f)$.  
Hence $C$ is an $f(X)$-certificate.

Secondly, note that for $1\leq i\leq b$ we must have
$|B_i|\leq bs_{X^{B_i}}(f)$: if we flip one of the $B_i$-variables 
in $X^{B_i}$ then the function value must flip from $f(X^{B_i})$ to $f(X)$ 
(otherwise $B_i$ would not be minimal), so every $B_i$-variable 
forms a sensitive set for $f$ on input $X^{B_i}$.
Hence the size of $C$ is $|\cup_i B_i|=\sum_{i=1}^b|B_i|
\leq \sum_{i=1}^b bs_{X^{B_i}}(f)\leq bs(f)^2$.
\end{proof}

The crucial lemma is the following, which we prove along the lines 
of~\cite[Lemma~4.1]{nisan:pram&dt}.

\begin{lemma}\label{lembounddf}
$D(f)\leq C^{(1)}(f)bs(f)$.
\end{lemma}

\begin{proof}
The following describes an algorithm to compute $f(X)$, querying at
most $C^{(1)}(f)bs(f)$ variables of $X$
(in the algorithm, by a ``consistent'' certificate $C$ or input $Y$ at 
some point we mean a $C$ or $Y$ that agrees with the values of all variables 
queried up to that point).
\begin{enumerate}
\item Repeat the following at most $bs(f)$ times:\\
Pick a consistent $1$-certificate $C$ and query those of its variables whose 
$X$-values are still unknown (if there is no such $C$, then return 0 and stop);
if the queried values agree with $C$ then return 1 and stop.
\item Pick a consistent $Y\in\{0,1\}^N$ and return $f(Y)$.
\end{enumerate}
The nondeterministic ``pick a $C$'' and ``pick a $Y$'' can easily be
made deterministic by choosing the first $C$ resp.\ $Y$ in some fixed order.
Call this algorithm $\bf A$.
Since $\bf A$ runs for at most $bs(f)$ stages and each stage queries at most 
$C^{(1)}(f)$ variables, $\bf A$ queries at most $C^{(1)}(f)bs(f)$ variables.

It remains to show that $\bf A$ always returns the right answer.
If it returns an answer in step~1, this is either because there are
no consistent $1$-certificates left (and hence $f(X)$ must be 0)
or because $X$ is found to agree with a particular $1$-certificate $C$;
in both cases $\bf A$ gives the right answer.

Now consider the case where $\bf A$ returns an answer in step~2.
We will show that all consistent $Y$ must have the same $f$-value.
Suppose not.
Then there are consistent $Y,Y'$ with $f(Y)=0$ and $f(Y')=1$.
$\bf A$ has queried $b=bs(f)$ $1$-certificates $C_1,C_2,\ldots,C_b$. 
Furthermore, $Y'$ contains a consistent $1$-certificate $C_{b+1}$.
We will derive from these $C_i$ disjoint sets $B_i$ such that $f$
is sensitive to each $B_i$ on $Y$.
For every $1\leq i\leq b+1$, define $B_i$ as the set of variables
on which $Y$ and $C_i$ disagree. 
Clearly, each $B_i$ is non-empty.
Note that $Y^{B_i}$ agrees with $C_i$, so $f(Y^{B_i})=1$ which shows that
$f$ is sensitive to each $B_i$ on $Y$.
Let $v$ be a variable in some $B_i$ ($1\leq i\leq b$),
then $X(v)=Y(v)\neq C_i(v)$.
Now for $j>i$, $C_j$ has been chosen consistent with all variables queried 
up to that point (including $v$), so we cannot have $X(v)=Y(v)\neq C_j(v)$,
hence $v\not\in B_j$. 
This shows that all $B_i$ and $B_j$ are disjoint.
But then $f$ is sensitive to $bs(f)+1$ disjoint sets on $Y$, 
which is a contradiction. 
Accordingly, all consistent $Y$ in step~2 must have the same $f$-value, 
and $\bf A$ returns the right value $f(Y)=f(X)$ in step~2,
because $X$ is one of those consistent $Y$.
\end{proof}

The inequality of the previous lemma is tight, because if $f=$ OR, 
then $D(f)=N$, $C^{(1)}(f)=1$, $bs(f)=N$.

The previous two lemmas imply $D(f)\leq bs(f)^3$.
Combining this with Theorem~\ref{thtbs} ($bs(f)\leq 16\ Q_2(f)^2$), 
we obtain the main result:

\begin{theorem}\label{thdoq6}
If $f$ is a Boolean function, then $D(f)\leq 4096\ Q_2(f)^6$.
\end{theorem}

We do not know if the $O(Q_2(f)^6)$-relation is tight, and suspect that it is not.
The best separation we know is for OR and similar functions, where
$D(f)=N$ and $Q_2(f)\in\Theta(\sqrt{N})$.
However, for such symmetric Boolean function we can do no better
than a quadratic separation: $D(f)\leq N$ always holds, 
and we have $Q_2(f)\in\Omega(\sqrt{N})$ by Theorem~\ref{thcountsym}, 
hence $D(f)\in O(Q_2(f)^2)$ for symmetric $f$.
For {\em monotone} Boolean functions, where the function value either 
increases or decreases monotonically if we set more input bits to 1, 
we can use \cite[Proposition~2.2]{nisan:pram&dt} ($bs(f)=C(f)$)
to prove $D(f)\leq 256\ Q_2(f)^4$. 
For the case of exact computation we can also give a better result:
Nisan and Smolensky (unpublished~\cite{nisan:commdeg4}) proved 
$D(f)\leq 2\ deg(f)^4$ for any $f$, which together with our 
Theorem~\ref{thexactpol} yields $D(f)\leq 32\ Q_E(f)^4$.

As a by-product, we improve the polynomial relation between $D(f)$ and 
$\widetilde{deg}(f)$.  
Nisan and Szegedy~\cite[Theorem~3.9]{nisan&szegedy:degree} proved 
$\widetilde{deg}(f)\leq D(f)\leq 1296\ \widetilde{deg}(f)^8.$            
Using our result $D(f)\leq bs(f)^3$ and Nisan and Szegedy's 
$bs(f)\leq 6\ \widetilde{deg}(f)^2$ \cite[Lemma~3.8]{nisan&szegedy:degree} 
we get

\begin{corollary} 
$\widetilde{deg}(f)\leq D(f)\leq 216\ \widetilde{deg}(f)^6.$               
\end{corollary}

\section{Some particular functions}\label{secpartfns}

First we will consider the OR-function, which is related to database search.
By Grover's well-known search algorithm~\cite{grover:search,bbht:bounds},
if at least one $x_i$ equals 1, we can find an index $i$ such that 
$x_i=1$ with high probability of success in $O(\sqrt{N})$ queries.
This implies that we can also compute the OR-function with high success
probability in $O(\sqrt{N})$: let Grover's algorithm generate an index $i$, 
and return $x_i$. 
Since $bs(\OR)=N$, Theorem~\ref{thtbs} gives us a lower bound of 
$\frac{1}{4}\sqrt{N}$ on computing the OR with bounded error probability,%
\footnote{This $\Omega(\sqrt{N})$ lower bound on search is actually quite 
well known~\cite{bbbv:str&weak,grover:search}, and is given in a tighter form 
in~\cite{bbht:bounds,zalka:grover}, but the way we obtained it here 
is rather different from existing proofs.}
so we have $Q_2(\OR)\in\Theta(\sqrt{N})$, where classically we require $\Theta(N)$
queries.
Now suppose we want to get rid of the probability of error:
can we compute the OR exactly or with zero-error using $O(\sqrt{N})$ queries?
If not, can quantum computation give us at least {\em some} advantage over
the classical deterministic case? 
Both questions have a negative answer:

\begin{proposition}\label{prorzero}
$Q_0(\OR)=N$.
\end{proposition}

\begin{proof}
Consider a network that computes OR with zero-error using $T=Q_0(\OR)$ queries.
By Lemma~\ref{lemamplpol}, there are complex-valued polynomials $p_k$ of degree 
at most $T$, such that the final state of the network on black-box $X$ is  
$$
\st{\phi^X}=\sum_{k\in K}p_k(X)\st{k}.
$$
Let $B$ be the set of all basis states ending in $10$ 
(i.e.,\ where the output is the answer 0).
Then for every $k\in B$ we must have $p_k(X)=0$ if $X\neq\vec{0}=(0,\ldots,0)$,
otherwise the probability of getting the incorrect answer $0$ on $\st{\phi^X}$ would be 
non-zero. On the other hand, there must be at least one $k'\in B$ such that 
$p_{k'}(\vec{0})\neq 0$, since the probability of getting the correct answer 0
on $\st{\phi^{\vec{0}}}$ must be non-zero. 
Let $p(X)$ be the real part of $1-p_{k'}(X)/p_{k'}(\vec{0})$.
This polynomial $p$ has degree at most $T$ and represents OR.
But then $p$ must have degree at least $deg(\OR)=N$, so $T\geq N$.
\end{proof}


\begin{corollary}\label{corexsearchn}
A quantum network for exact or zero-error search requires $N$ queries.
\end{corollary}

Under the promise that the number of solutions is either 0 or $K$, 
for some fixed known $K$, exact search can be done in $O(\sqrt{N/K})$ 
queries~\cite{hoyer:comm,mosca:eigen}.
A partial block sensitivity argument (see the comment following Theorem~\ref{thtbs})
shows that this is optimal up to a multiplicative constant.

Like the OR-function, PARITY has $deg(f)=N$, so by Theorem~\ref{thexactpol} 
exact computation requires at least $N/2$ queries. 
This is also sufficient.
It is well known that the XOR of 2 variables can be computed using only one
query~\cite{cemm:revisited}.
We can group the $N$ variables of $X$ as $N/2$ pairs: 
$(x_0,x_1),(x_2,x_3),\ldots,(x_{N-2},x_{N-1})$,
and compute the XOR of all $N/2$ pairs using $N/2$ queries. 
The parity of $X$ is the parity of these $N/2$ XOR values,
which can be computed without any further queries.
If we allow bounded-error, then $N/2$ queries of course still suffice. 
It follows from Theorem~\ref{thapprpol} that this cannot be improved, because
$\widetilde{deg}(\mbox{\rm PARITY})=N$~\cite{minsky&papert:perceptrons}:

\begin{lemma}[Minsky, Papert]\label{lemapproxparity}
$\widetilde{deg}(\mbox{\rm PARITY})=N$.
\end{lemma}

\begin{proof}
Let $f$ be PARITY on $N$ variables.
Let $p$ be a polynomial of degree $\widetilde{deg}(f)$ that approximates $f$.
Since $p$ approximates $f$, its symmetrization $p^{sym}$ also approximates $f$.
By Lemma~\ref{lemsym}, there is a polynomial $q$, of degree at most
$\widetilde{deg}(f)$, such that $q(|X|)=p^{sym}(X)$ for all inputs.
Thus we must have $|f(X)-q(|X|)|\leq 1/3$, so
\begin{quote}
$q(0)\leq 1/3$, $q(1)\geq 2/3$, \ldots,
$q(N-1)\geq 2/3$, $q(N)\leq 1/3$ (assuming $N$ even).
\end{quote}
We see that the polynomial $q(x)-1/2$ must have at least $N$ zeroes,
hence $q$ has degree at least $N$ and $\widetilde{deg}(f)\geq N$.
\end{proof}

\begin{proposition}
If $f$ is PARITY on $\{0,1\}^N$, then $Q_E(f)=Q_0(f)=Q_2(f)=N/2$.%
\footnote{Recently, this has also been proved by Farhi, Goldstone, 
Gutmann, and Sipser~\cite{fggs:parity}, using a different technique.
As noted independently by Terhal~\cite{terhal:comm} and 
\cite{fggs:parity}, this result immediately implies results by 
Ozhigov~\cite{ozhigov:box} to the effect that no quantum computer
can significantly speed up the computation of {\em all} functions 
(this follows because no quantum computer can significantly speed 
up the computation of PARITY).}
\end{proposition}
 
For {\em classical} deterministic or randomized methods, 
$N$ queries are necessary in both the exact and the zero-error setting.
($R(\mbox{PARITY})=\ceil{N/3}$ because for $R(f)$ we count {\em expected} 
number of queries.)
Note that while computing PARITY on a quantum network is much harder 
than OR in the {\em bounded-error} setting ($N/2$ versus 
$\Theta(\sqrt{N})$), in the {\em exact} setting PARITY is actually easier 
($N/2$ versus $N$).

The upper bound on PARITY uses the fact that the XOR connective can be computed
with only one query. Using polynomial arguments, it turns out that XOR and 
its negation are the {\em only} examples among all $16$ connectives where quantum gives 
an advantage over classical computation.

Since the AND of $N$ variables can be reduced 
to MAJORITY on $2N-1$ variables (if we set the first $N-1$ variables to 0,
then the MAJORITY of all variables equals the AND of the last $N$ variables) 
and AND, like OR, requires $N$ queries to be computed exactly or with 
zero-error, MAJORITY takes at least $(N+1)/2$ queries.
Van Melkebeek~\cite{melkebeek:comm} and Hayes and Kutin
\cite{hayes&kutin:comm} independently found an exact quantum algorithm that uses
at most $N+1-e(N)$ queries, where $e(N)$ is the number of 1s in the
binary representation of $N$; this can save up to $\log N$ queries.
For the zero-error case, the $(N+1)/2$ lower bound applies;
Van Melkebeek, Hayes and Kutin have found an algorithm that works in
roughly $\sqrt{0.5}N$ queries.
For the bounded-error case, we can apply Theorem~\ref{thcountsym}:
if $f=$ MAJORITY, then $\Gamma(f)=1$, so we need $\Theta(N)$ queries.
The best upper bound we have here is $N/2+\sqrt{N}$, 
which follows from~\cite{dam:oracle}.

\section*{Acknowledgments}
We would like to thank Lance Fortnow for stimulating discussions on many of 
the topics treated here; Alain Tapp for sending us a preliminary version 
of~\cite{bht:counting} and subsequent discussions about quantum counting;
Andris Ambainis for sending us his proof that most functions cannot 
be computed with bounded-error using significantly fewer than $N$ queries;
Noam Nisan for sending us his proof that $D(f)\leq 2\ deg(f)^4$;
Dieter van Melkebeek, Tom Hayes, and Sandy Kutin for their algorithms 
for MAJORITY; and Hayes and Kutin for the reference to~\cite{gathen&roche:poly}.
R.C.\ and M.M.\ gratefully acknowledge the hospitality of the CWI, where 
much of this research took place.
M.M.\ thanks CESG for their support.


\begin{thebibliography}{10}\setlength{\itemsep}{-1ex}\small

\bibitem{ambainis:comm}
A.~Ambainis.
\newblock Personal communication, February 1998.

\bibitem{beigel:poly}
R.~Beigel.
\newblock The polynomial method in circuit complexity.
\newblock In {\em Proceedings of the 8th {IEEE} Structure in Complexity Theory
  Conference}, pages 82--95, 1993.

\bibitem{bbbv:str&weak}
C.~H. Bennett, E.~Bernstein, G.~Brassard, and U.~Vazirani.
\newblock Strengths and weaknesses of quantum computing.
\newblock {\em SIAM Journal of Computing}, 26(5):1510--1523, 1997.
\newblock Available at http://xxx.lanl.gov/abs/quant-ph/9701001.

\bibitem{boneh&lipton}
D.~Boneh and R.~J. Lipton.
\newblock Quantum cryptanalysis of hidden linear functions (extended abstract).
\newblock In {\em Advances in Cryptology (CRYPTO'95)}, volume 963 of {\em
  Lecture Notes in Computer Science}, pages 424--437. Springer, 1995.

\bibitem{bbht:bounds}
M.~Boyer, G.~Brassard, P.~H{\o}yer, and A.~Tapp.
\newblock Tight bounds on quantum searching.
\newblock {\em Fortschritte der Physik}, 46(4--5):493--505, 1998.
\newblock Earlier version in Physcomp'96; also quant-ph/9605034.

\bibitem{brassard&hoyer:simon}
G.~Brassard and P.~H{\o}yer.
\newblock An exact quantum polynomial-time algorithm for {S}imon's problem.
\newblock In {\em Proceedings of the 5th Israeli Symposium on Theory of
  Computing and Systems (ISTCS'97)}, pages 12--23, 1997.
\newblock quant-ph/9704027.

\bibitem{bht:counting}
G.~Brassard, P.~H{\o}yer, and A.~Tapp.
\newblock Quantum counting.
\newblock In {\em Proceedings of 25th ICALP}, volume 1443 of {\em Lecture Notes
  in Computer Science}, pages 820--831. Springer, 1998.
\newblock quant-ph/9805082.

\bibitem{BuhrmanCleveWigderson98}
H.~Buhrman, R.~Cleve, and A.~Wigderson.
\newblock Quantum vs.\ classical communication and computation (preliminary
  version).
\newblock In {\em Proceedings of 30th STOC}, pages 63--68, 1998.
\newblock quant-ph/9802040.

\bibitem{cemm:revisited}
R.~Cleve, A.~Ekert, C.~Macchiavello, and M.~Mosca.
\newblock Quantum algorithms revisited.
\newblock In {\em Proceedings of the Royal Society of London}, volume A454,
  pages 339--354, 1998.
\newblock quant-ph/9708016.

\bibitem{deutsch&jozsa}
D.~Deutsch and R.~Jozsa.
\newblock Rapid solution of problems by quantum computation.
\newblock In {\em Proceedings of the Royal Society of London}, volume A439,
  pages 553--558, 1992.

\bibitem{ehlich&zeller:schwankung}
H.~Ehlich and K.~Zeller.
\newblock Schwankung von {P}olynomen zwischen {G}itterpunkten.
\newblock {\em Mathematische Zeitschrift}, 86:41--44, 1964.

\bibitem{fggs:parity}
E.~Farhi, J.~Goldstone, S.~Gutmann, and M.~Sipser.
\newblock A limit on the speed of quantum computation in determining parity.
\newblock quant-ph/9802045, 16 Feb 1998.

\bibitem{ffkl:toolkit}
S.~Fenner, L.~Fortnow, S.~Kurtz, and L.~Li.
\newblock An oracle builder's toolkit.
\newblock In {\em Proceedings of the 8th {IEEE} Structure in Complexity Theory
  Conference}, pages 120--131, 1993.

\bibitem{fortnow&rogers:limitations}
L.~Fortnow and J.~Rogers.
\newblock Complexity limitations on quantum computation.
\newblock In {\em Proceedings of the 13th {IEEE} Conference on Computational
  Complexity}, pages 202--209, 1998.

\bibitem{grover:search}
L.~K. Grover.
\newblock A fast quantum mechanical algorithm for database search.
\newblock In {\em Proceedings of 28th STOC}, pages 212--219, 1996.

\bibitem{hayes&kutin:comm}
T.~Hayes and S.~Kutin.
\newblock Personal communication, May 1998.

\bibitem{hoyer:conjugated}
P.~H{\o}yer.
\newblock Conjugated operators in quantum algorithms.
\newblock Preprint, 1997.

\bibitem{hoyer:comm}
P.~H{\o}yer.
\newblock Personal communication, January 1998.

\bibitem{kitaev:stabilizer}
A.~Y. Kitaev.
\newblock Quantum measurements and the {A}belian stabilizer problem.
\newblock quant-ph/9511026, 12 Nov 1995.

\bibitem{minsky&papert:perceptrons}
M.~Minsky and S.~Papert.
\newblock {\em Perceptrons}.
\newblock MIT Press, Cambridge, MA, 1968.
\newblock Second, expanded edition 1988.

\bibitem{mosca:eigen}
M.~Mosca.
\newblock Quantum searching, counting and amplitude amplification by
  eigenvector analysis.
\newblock In {\em MFCS'98 workshop on Randomized Algorithms}, 1998.

\bibitem{mosca&ekert:hidden}
M.~Mosca and A.~Ekert.
\newblock The hidden subgroup problem and eigenvalue estimation on a quantum
  computer.
\newblock In {\em Proceedings of NASA QCQC conference}, volume 1509 of {\em
  Lecture Notes in Computer Science}. Springer, 1998.

\bibitem{nisan:pram&dt}
N.~Nisan.
\newblock {CREW PRAM}s and decision trees.
\newblock {\em SIAM Journal of Computing}, 20(6):999--1007, 1991.
\newblock Earlier version in STOC'89.

\bibitem{nisan:commdeg4}
N.~Nisan.
\newblock Personal communication, June 1998.

\bibitem{nisan&szegedy:degree}
N.~Nisan and M.~Szegedy.
\newblock On the degree of {B}oolean functions as real polynomials.
\newblock {\em Computational Complexity}, 4(4):301--313, 1994.
\newblock Earlier version in STOC'92.

\bibitem{ozhigov:box}
Y.~Ozhigov.
\newblock Quantum computer can not speed up iterated applications of a black
  box.
\newblock quant-ph/9712051, 22 Dec 1997.

\bibitem{paturi:degree}
R.~Paturi.
\newblock On the degree of polynomials that approximate symmetric {B}oolean
  functions (preliminary version).
\newblock In {\em Proceedings of 24th STOC}, pages 468--474, 1992.

\bibitem{rivlin&cheney:approx}
T.~J. Rivlin and E.~W. Cheney.
\newblock A comparison of uniform approximations on an interval and a finite
  subset thereof.
\newblock {\em SIAM Journal of Numerical Analysis}, 3(2):311--320, 1966.

\bibitem{saks&wigderson:trees}
M.~Saks and A.~Wigderson.
\newblock Probabilistic {B}oolean decision trees and the complexity of
  evaluating game trees.
\newblock In {\em Proceedings of 27th FOCS}, pages 29--38, 1986.

\bibitem{santha:montecarlo}
M.~Santha.
\newblock On the {M}onte {C}arlo decision tree complexity of read-once
  formulae.
\newblock In {\em Proceedings of the 6th {IEEE} Structure in Complexity Theory
  Conference}, pages 180--187, 1991.

\bibitem{shor:factoring}
P.~W. Shor.
\newblock Polynomial-time algorithms for prime factorization and discrete
  logarithms on a quantum computer.
\newblock {\em SIAM Journal of Computing}, 26(5):1484--1509, 1997.
\newblock Earlier version in FOCS'94; also quant-ph/9508027.

\bibitem{simon:power}
D.~Simon.
\newblock On the power of quantum computation.
\newblock {\em SIAM Journal of Computing}, 26(5):1474--1483, 1997.
\newblock Earlier version in FOCS'94.

\bibitem{terhal:comm}
B.~Terhal.
\newblock Personal communication, December 1997.

\bibitem{dam:oracle}
W.~van Dam.
\newblock Quantum oracle interrogation: Getting all information for almost half
  the price.
\newblock In {\em Proceedings of 39th FOCS}, 1998.
\newblock quant-ph/9805006.

\bibitem{melkebeek:comm}
D.~van Melkebeek.
\newblock Personal communication, May 1998.

\bibitem{gathen&roche:poly}
J.~von~zur Gathen and J.~R. Roche.
\newblock Polynomials with two values.
\newblock {\em Combinatorica}, 17(3):345--362, 1997.

\bibitem{zalka:grover}
C.~Zalka.
\newblock Grover's quantum searching algorithm is optimal.
\newblock quant-ph/9711070, 26 Nov 1997.

\end{thebibliography}

\end{document}